# The (Elementary) Mathematical Data Model Revisited

**Christian Mancas***

*Mathematics and Computer Science Dept., Ovidius University at Constanta, Romania*

***Corresponding Author:** Christian Mancas, Mathematics and Computer Science Dept., Ovidius University at Constanta, Romania.

**Abstract**

This paper presents the current version of our (Elementary) Mathematical Data Model ((E)MDM), which is based on the naïve theory of sets, relations, and functions, as well as on the first-order predicate calculus with equality. Many real-life examples illustrate its 4 types of sets, 4 types of functions, and 76 types of constraints. This rich panoply of constraints is the main strength of this model, guaranteeing that any data value stored in a database is plausible, which is the highest possible level of syntactical data quality. An (E)MDM example scheme is presented and contrasted with some popular family tree software products.

## Introduction

The (Elementary) Mathematical Data Model ((E)MDM) was introduced first in Romania [1-4] and then internationally in [5-7]. Its main advantage is its plethora of constraint types that grew continuously every year, whenever a new real-life example was encountered. The last published paper on (E)MDM [8], considered only 60 explicit constraint types; currently, there are 76.

Data quality is paramount for any software system: if you let a database (db) store unplausible data ("garbage in"), then both information and knowledge computed on that data will be unplausible ("garbage out"). Software systems cannot decide whether a data value is correct or not: for example, most probably, only a handful of people know exactly what was HM Queen Elizabeth II height on her death bed; how could a software application know it? But any software application must accept for db storing only plausible values: for example, for current humans, height must be between 40cm (otherwise chances of survival are almost zero) and 275cm (Robert Wadlaw, the tallest man ever, had 272cm).

Constraints are formalizing business rules. If you do not enforce a single such rule, your db might store unplausible values. Dually, of course, if you add to your software system rules that do not exist in the corresponding subuniverse of discourse, you prevent storing plausible data values in the db. This is why discovering and enforcing all business rules governing the subuniverse you model is crucial.

Classifying constraints in mathematical types is not only beneficial to data modeling but also allows for their algorithmic enforcement through automatically generated software code. Thus, (E)MDM is also a modelware tool, i.e., a member of the family of the 5th generation of programming languages [9].

Database theory is pure applied naïve theory of sets, relations, and functions, as well as of the first-order predicate logic with equality (which formalizes both constraints, as closed clauses, and queries, as open ones). As this is almost entirely taught during K12 studies, we've added (Elementary) to the name of our model.

We introduced (E)MDM as a cornerstone modelling level between the Entity-Relationship Data Models (E-RDM) and Relational Data Model (RDM) schemata [10], for both validating and refining the E-RDM models before translating them into RDM schemata and sets of non-relational constraints that must be enforced by the software applications managing the corresponding dbs.

The following section presents the current version of the (E)MDM. Section 3 presents an example of (E)MDM modeling. Section 4 discusses it against some popular family tree software products. Section 5 abstracts related work. The paper ends with conclusion and a reference list.

## The current version of (E)MDM

A(n) (E)MDM db scheme is a quadruple $\mathcal{DKS} = < S, \mathcal{M}, C; \mathcal{P} >$, where $S$ is a finite non-empty poset by inclusion, $\mathcal{M}$ is a finite non-empty set of mappings defined on and taking values from the sets of $S$, $C$ is a finite non-empty set of constraints (i.e., closed Horn clauses of the first-order predicate logic with equality) over sets in $S$ and/or mappings in $\mathcal{M}$, and $\mathcal{P}$ is a finite set of Datalog¬ programs also over sets in $S$ and mappings in $\mathcal{M}$. Whenever $\mathcal{P}$ is empty, $\mathcal{DKS}$ is a db scheme, otherwise it is a knowledge base one.

### The poset of sets

$(S, \subseteq)$ is a poset of sets, with $S = \Omega \oplus \mathcal{V} \oplus {}^*S \oplus SysS$, where:

- $\Omega = \mathcal{E} \oplus \mathcal{R}$ (the non-empty collection of *object sets*), where:
  * $\mathcal{E}$ is a non-empty collection of atomic *entity-type sets*, e.g., PEOPLE, COUNTRIES, PRODUCTS.
  * $\mathcal{R}$ is a collection of *relationship-type sets*, e.g., NEIGHBOUR_COUNTRIES $\subset$ COUNTRIES x COUNTRIES, STOCKS $\subset$ PRODUCTS x WAREHOUSES.
- $\mathcal{V}$ is a non-empty collection of *value sets*, e.g., R_C = {"r", "o", "y", "g", "b", "i", "v"}, SEXES = {"F","M"}, [0,16] $\subset$ NAT(2), ASCII(32) $\subset$ ASCII($n$), CURRENCY(8) $\subset$ RAT(10,2), [1/1/100, *Today*()] $\subset$ DATETIME (with NAT($n$) being the subset of naturals of at most $n$ digits, RAT($n, m$) the subset of rationals of at most $n$ digits before the decimal point and $m$ after it, ASCII($n$) the subset of the freely generated monoid over the ASCII alphabet only including strings of maximum length $n$, etc.).
- ${}^*S$ is a collection of *computed sets*, e.g., MALES, FEMALES, UNPAID_BILLS, FREE_COUNTRIES.
- $SysS$ is a collection of *system sets*, e.g., $\varnothing$ (the empty set), NULLS (the distinguished countable set of null values), BOOLE = {*true*, *false*}, NAT($n$), RAT($n, m$), ASCII($n$), DATETIME (the set of date and time values).

In RDM schemata, generally, the object sets are tables, the computed sets are views (queries), the system sets are corresponding db management system (RDBMS) data types, and the value sets are their needed subsets.

### The set of mappings

$\mathcal{M} = \mathcal{A} \oplus \mathcal{F} \oplus {}^*\mathcal{M} \oplus SysM$, where:

- $\mathcal{A} \subset \text{Hom}(S\text{-}SysS \oplus \mathcal{V}, \mathcal{V})$ is the non-empty *set of attributes*, e.g., $x$ : PEOPLE $\leftrightarrow$ NAT(10), *FirstName* : PEOPLE $\rightarrow$ ASCII(64), *Birth-Date* : PEOPLE $\rightarrow$ [1/1/-6000, Today()], *Sex* : PEOPLE $\rightarrow$ {"F", "M"}, $x$ : COUNTRIES $\leftrightarrow$ NAT(3), *CountryName* : COUNTRIES $\rightarrow$ ASCII(128), *TelPrefix* : COUNTRIES $\rightarrow$ NAT(4), *Stock* : STOCKS $\rightarrow$ [0, 100], *Amount* : UNPAID_BILLS $\rightarrow$ (0, 100000], where $\leftrightarrow$ denotes injections (one-to-one functions) and $x$ is our notation for *object identifiers*, i.e., functions to be implemented as autonumber surrogate primary keys.

- $\mathcal{F} \subset$ Hom($S\text{-}SysS \oplus \mathcal{V}$, $S\text{-}SysS \oplus \mathcal{V}$) is the non-empty set of *structural functions*, e.g., *BirthPlace* : PEOPLE → CITIES, *Capital* : COUNTRIES ↔ CITIES, *Country* : CITIES → COUNTRIES, *Representative* : DISTRICTS ↔ PEOPLE, *Customer* : UNPAID_BILLS → CUSTOMERS.
- $*\mathcal{M}$ is the set of *computed mappings*, e.g., *BirthCountry* = *Country* ° *BirthPlace* : PEOPLE → COUNTRIES, *Mother* • *Father* : PEOPLE → PEOPLE x PEOPLE, *Senator1* • *Senator2* : STATES → PEOPLE x PEOPLE, *Age* = *Years(Today()-BirthDate)* : PEOPLE → [0, 175].
- $SysM$ is the set of *system mappings*, e.g., $\mathbf{1}_S$ (the unity function of a set *S*), *card*(*S*) (the cardinal of a set *S*), *Im(f), PreIm(f), Ker(f)* (the image, pre-image, and kernel of a function *f*), *Len, Lft, Rgt, Mid* (the length, left, right, and middle parts of a string), etc.

In RDM schemata, the attributes, structural functions, and computed mappings are implemented as table or/and view (query) columns, with structural functions being foreign keys.

**The set of constraints**

$C = SC \oplus \mathcal{MC} \oplus \mathcal{OC} \oplus SysC$, where:

**Set constraints**

There are two blocks of set constraints: general and dyadic relation ones.

- $SC = SGC \oplus SDRC$ is the set of *set constraints*, where:
  ✓ $SGC = SIC \oplus SEC \oplus SDC \oplus SUC \oplus SDSC$ is the set of *general set constraints*, where:
    ❖ $SIC$ is the set of *inclusion constraints*, e.g., DRIVERS ⊆ EMPLOYEES ⊆ PEOPLE ⊆ CUSTOMERS, PREEQUISITES ⊆ COURSES.
    ❖ $SEC$ is the set of *set equality constraints*, e.g., TAKEOFF_AIRPORTS = AIRPORTS, LANDING_AIRPORTS = AIRPORTS.
    ❖ $SDC$ is the set of *disjointness constraints*, e.g., ELECTED_OFFICIALS ∩ STATE_CONTRACTORS = ∅, INCOMPATIBLE_DRUGS ∩ OTHER_DRUGS = ∅.
    ❖ $SUC$ is the set of *union constraints*, e.g., INCOMPATIBLE_DRUGS = AMINOGLYCOSIDES ∪ CHLORDIAZEPOXIDE ∪ DIAZEPAM ∪ DIGITALIS_ GLYCOSIDES ∪ PENTOBARBITAL ∪ PHENYTOIN ∪ SECOBARBITAL ∪ SODIUM_BICARBONATE ∪ THEOPHYLLINE_DERIVATIVES.
    ❖ $SDSC$ is the set of *direct sum constraints*, e.g., MPS ⊆ COMMONERS ⊕ LORDS, CONGRESSMEN = REPRESENTATIVES ⊕ SENATORS.

Among these 5 constraint types, only inclusion is fundamental: equalities are double inclusions, and the 3 remaining ones are particular cases of equality.

  ✓ $SDRC = DRRC \oplus DRIRC \oplus DRSC \oplus DRASC \oplus DRTC \oplus DRITC \oplus DREC$.

$DRIEC \oplus DRQC \oplus DRAC \oplus DRCC$ is the set of *dyadic relation constraints*, where:

  ❖ $DRRC$ is the set of *dyadic relation reflexivity constraints*, e.g., *HasSameColorAs, HasSameSizeAs, HasSameShapeAs, IsAtLeast-AsComplicatedAs, LivesInSameCityAs, IsBloodRelatedTo, WeighsNo-MoreThan.*
  ❖ $DRIRC$ is the set of *dyadic relation irreflexivity constraints*, e.g., *IsInFrontOf, OccuredEarlier(Later)Than, IsAdjoinTo, IsLargerThan, IsSmallerThan, IsLeftOf, IsRightOf, Prerequisites, AirportConnections, Distances.*
  ❖ $DRSC$ is the set of *dyadic relation symmetry constraints*, e.g., : *HasSameSizeAs, HasSameShapeAs, IsAdjoinTo, HasSameColorAs, LivesInSameCityAs, IsBloodRelatedTo, IsSiblingOf, WeighsNoMoreThan, IsAtLeastAsComplicatedAs, IsLocatedWithinXmOf, IsMarriedTo, WentOnDateWith, AirportConnections.*
  ❖ $DRASC$ is the set of *dyadic relation asymmetry constraints*, e.g., *IsFatherOf, IsMotherOf, IsChildOf, IsYounger(Older)Than, OccuredEarlier(Later)Than, WeighsMoreThan, IsLargerThan, IsSmallerThan, IsLeftOf, IsRightOf, IsInFrontOf, Prerequisites, Distances.*

- ❖ $\mathcal{DRTC}$ is the set of *dyadic relation transitivity constraints*, e.g., *IsYounger(Older)Than, OccurredEarlier(Later)Than, Weighs(No)MoreThan, IsLargerThan, IsSmallerThan, IsBloodRelatedTo, IsAtLeastAsComplicatedAs, IsInFrontOf, HasArrivedSooner(Later)Than, IsPoorer(Richer)Than, IsAnAncestor(Descendant)Of, HasSameShapeAs, HasSameSizeAs, HasSameColorAs, LivesInSameCityAs, IsDividing, IsLeftOf, IsRightOf, IsAncestorOf, IsDescendantOf, AirportConnections, Distances.*
- ❖ $\mathcal{DRITC}$ is the set of *dyadic relation intransitivity constraints*, e.g., *IsMotherOf, IsFatherOf, IsChildrenOf, Prerequisites.*
- ❖ $\mathcal{DREC}$ is the set of *dyadic relation Euclideanity constraints*, e.g., *IsBloodRelatedTo, HasSameShapeAs, HasSameSizeAs, HasSameColorAs, LivesInSameCityAs, IsSiblingOf* (in (E)MDM, a Euclidean dyadic relation is both left and right Euclidean).
- ❖ $\mathcal{DRIEC}$ is the set of *dyadic relation inEuclideanity constraints*, e.g., *IsMotherOf, IsFatherOf, IsChildrenOf, IsAncestor* (in (E)MDM, an inEuclidean dyadic relation is neither left, nor right Euclidean).
- ❖ $\mathcal{DRQC}$ is the set of *dyadic relation equivalence constraints*, e.g., *IsAtLeastAsComplicatedAs, HasSameCitizenshipAs, HasSame(Time)Length-As, HasSameReligionAs, HasSameChildrenAs, HasSameEmployersAs, Has-Same(Time)LengthAs, WeighsNoMoreThan, HasSameColorAs, HasSame-SizeAs, HasSameRaceAs, HasSameShapeAs, LivesInSameCityAs, Graduated-SameSchool(Alumni)As, IsBloodRelatedTo.*
- ❖ $\mathcal{DRAC}$ is the set of *dyadic relation acyclicity constraints*, e.g., *IsFatherOf, IsMotherOf, IsChildOf, IsYounger(Older)Than, OccuredEarlier(Later)Than, WeighsMoreThan, IsLargerThan, IsSmallerThan, IsLeftOf, IsRightOf, IsInFrontOf, Prerequisites.*
- ❖ $\mathcal{DRCC}$ is the set of *dyadic relation connectivity constraints*, e.g., *BelongsTo-SameGroupAs*, no matter what group is involved, from the algebraic to social media ones (in (E)MDM, connectivity is weak, i.e., it includes $x \neq y$, for any pair <$x, y$> of connected elements, which is also called by some authors *connex* or *completeness*).

None of these 11 constraint types is fundamental, as dyadic relations are particular cases of homogeneous binary function products (see the corresponding block from $\mathcal{HBFPC}$).

The grand total of set constraint types is 16, with only 1 fundamental.

*Mapping constraints*

There are five blocks of mapping constraints: general, self-map, function product, homogeneous binary function products, and function diagram cycle ones.

- ➢ $\mathcal{MC}$ = $\mathcal{MGC} \oplus \mathcal{MSMC} \oplus \mathcal{MPC} \oplus \mathcal{HBFPC} \oplus \mathcal{FDCC}$ is the set of *mapping constraints*, where:
  - ✓ $\mathcal{MGC}$ = $\mathcal{MTC} \oplus \mathcal{MIC} \oplus \mathcal{MNPC} \oplus \mathcal{MSC} \oplus \mathcal{MBC} \oplus \mathcal{MDVC}$ is the set of *general mapping constraints*, where:
    - ❖ $\mathcal{MTC}$ is the set of *totality constraints*, e.g., $x$ : PEOPLE ↔ NAT(10) total, *FirstName* : PEOPLE ↔ ASCII(64) total, $x$ : COUNTRIES ↔ NAT(3) total, *Country-Name* : COUNTRIES → ASCII(128) total, *Country* : CITIES → COUNTRIES total (in (E)MDM, $f : D → C$ is total iff $C \cap$ NULLS = $\varnothing$).
    - ❖ $\mathcal{MIC}$ is the set of *single key (injectivity) constraints*, e.g., $x$ : PEOPLE ↔ NAT(10), $x$ : COUNTRIES ↔ NAT(3), *Capital* : COUNTRIES ↔ CITIES, *Representative* : DISTRICTS ↔ PEOPLE.
    - ❖ $\mathcal{MNPC}$ is the set of *non-primeness constraints*, e.g., *Stock* : STOCKS → [0, 100], *Amount* : UNPAID_ BILLS → (0, 100000], *Area* : COUNTRIES → NAT(8) (in (E)MDM, $f : D → C$ is *non-prime* iff, semantically, it cannot be either one-to-one or a member of a minimally one-to-one function product).
    - ❖ $\mathcal{MSC}$ is the set of *surjectivity constraints*, e.g., *Edition* : VOLUMES → EDITIONS, *County* : CITIES → COUNTIES.
    - ❖ $\mathcal{MBC}$ is the set of *bijectivity constraints*, e.g., *District* : REPRESENTATIVES ↔ DISTRICTS.
    - ❖ $\mathcal{MDVC}$ is the set of *default value constraints*, e.g., *Stock* : STOCKS → [0, 100] default = 0.

In total, there are 6 types of general mapping constraints, out of which only one-to-oneness (injectivity), non-primeness, ontoness (surjectivity), and default ones are fundamental: totality is a particular case of existence constraints (i.e., of type $\varnothing \vdash g$, see the mapping product block) and bijectivity derives from injectivity and surjectivity.

- ✓ $\mathcal{MPC} = \mathcal{MMIC} \oplus \mathcal{MSKC} \oplus \mathcal{MEC} \oplus \mathcal{MNEC}$ is the set of *general mapping product constraints*, where:
  - ❖ $\mathcal{MMIC}$ is the set of *concatenated key (minimal injectivity) constraints*, e.g., *StateName • Country* : *STATES* ↔ ASCII(64) x *COUNTRIES, Prerequisite • Course* : *PREREQUISITES* ↔ *COURSES* x *COURSES, City1 • City2* : *DISTANCES* ↔ *CITIES* x *CITIES*.
  - ❖ $\mathcal{MSKC}$ is the set of *subkey (variable geometry key) constraints*, e.g., *Folder • FName subkey of Folder • FName • FExt* : *FILES* → *FILES* x ASCII(255) x (ASCII(255) ∪ NULLS), whenever *FExt*(*x*) ∈ NULLS.
  - ❖ $\mathcal{MEC}$ is the set of *existence constraints*, e.g., *e-mail* ⊢— *Fname • Phone, Fname • Phone* ⊢— *e-mail* (in (E)MDM, just like in RDM, $f \vdash\!\!-\, g$ is a shorthand for $(\forall x)(f(x) \notin \text{NULLS} \Rightarrow g(x) \notin \text{NULLS})$; the only difference is that both *f* and *g* may be computed functions, by composition or/and Cartesian function product).
  - ❖ $\mathcal{MNEC}$ is the set of *non-existence constraints*, e.g., *TributaryTo* ¬⊢— *Lake • Sea • Ocean • LostInto*, where *TributaryTo* : *RIVERS* → *RIVERS, Lake* : *RIVERS* → *LAKES, Sea* : *RIVERS* → *SEAS, Ocean* : *RIVERS* → *OCEANS*, and *LostInto* : *RIVERS* → *GEOGRAPHIC_UNITS* (in (E)MDM, $f \neg \vdash\!\!-\, g$ is a shorthand for $(\forall x)(f(x) \notin \text{NULLS} \Rightarrow g(x) \in \text{NULLS})$).

In total, there are 4 types of general mapping product constraints, all of them being fundamental.

- ✓ $\mathcal{HBFPC} = \mathcal{HBFPCC} \oplus \mathcal{HBFPRC} \oplus \mathcal{HBFPNRC} \oplus \mathcal{HBFPNIC} \oplus \mathcal{HBFPIRC} \oplus \mathcal{HBFPSC} \oplus \mathcal{HBFPNSC} \oplus \mathcal{HBFPASC} \oplus \mathcal{HBFPTC} \oplus \mathcal{HBFPNTC} \oplus \mathcal{HBFPITC} \oplus \mathcal{HBFPEC} \oplus \mathcal{HBFPNEC} \oplus \mathcal{HBFPIEC} \oplus \mathcal{HBFPQC} \oplus \mathcal{HBFPNQC} \oplus \mathcal{HBFPACC}$ is the set of *homogeneous binary function product (hbfp) constraints* (i.e., of type $f • g : D \rightarrow C^2$; dyadic relations are particular cases of hbfps, for which $C \cap \text{NULLS} = \emptyset$, $f • g$ is minimally one-to-one, and *f* and *g* are called *canonical projections*), where:
  - ❖ $\mathcal{HBFPCC}$ is the set of *hbfp connectivity constraints*, e.g., : *Host • Visitor* : *CHAMPIONSHIP_MATCHES* → *TEAMS* x *TEAMS*.
  - ❖ $\mathcal{HBFPRC}$ is the set of *hbfp reflexivity constraints*, e.g., *Domain • Codomain* : $\mathcal{F} \rightarrow S$ x $S$, as, for any set $S \in \mathcal{S}$, there is a corresponding system unity mapping $\mathbf{1}_S : S \rightarrow S$, $\mathbf{1}_S(x) = x$, $\forall x \in S$.
  - ❖ $\mathcal{HBFPNRC}$ is the set of *hbfp null-reflexivity constraints*; in (E)MDM, given any dyadic relation constraint type *DRCT*, the null-DRCT corresponding constraint type requires that DRTC be satisfied for any not null-values; e.g., $f • g : D \rightarrow (C \cup \text{NULLS})^2$ is *null-reflexive* iff $(\forall x \in C)(Im(f • g) \supset (x, x) \vee Im(f • g) \supset (x, ) \vee Im(f • g) \supset (, x) \vee Im(f • g) \supset (, ))$.
  - ❖ $\mathcal{HBFPNIC}$ is the set of *hbfp null-identity constraints*, e.g., *BirthPlace • (City ° BirthClinic)* : *PEOPLE* → *CITIES* x *CITIES* (identity ones are useless: why storing in a same db table two columns whose values must be equal on each line?).
  - ❖ $\mathcal{HBFPIRC}$ is the set of *hbfp irreflexivity constraints*, e.g., *Host • Visitor* : *MATCHES* → *TEAMS* x *TEAMS*.
  - ❖ $\mathcal{HBFPSC}$ is the set of *hbfp symmetry constraints*, e.g., *TakeOffAirport • LandingAirport* : *AIRPORT_CONNECTIONS* → *AIRPORTS* x *AIRPORTS*.
  - ❖ $\mathcal{HBFPNSC}$ is the set of *hbfp null-symmetry constraints* (i.e., $(\forall x,y \in C)(Im(f • g) \supset (x, y) \Rightarrow Im(f • g) \supset (y, x) \vee Im(f • g) \supset (y, ) \vee Im(f • g) \supset (, x) \vee Im(f • g) \supset (, ))$.
  - ❖ $\mathcal{HBFPASC}$ is the set of *hbfp asymmetry constraints*, e.g., *Host • Visitor* : *ELIMINATORY_MATCHES* → *TEAMS* x *TEAMS*.
  - ❖ $\mathcal{HBFPTC}$ is the set of *hbfp transitivity constraints*, e.g., *TakeOffAirport • LandingAirport* : *AIRPORT_CONNECTIONS* → *AIRPORTS* x AIRPORTS.
  - ❖ $\mathcal{HBFPNTC}$ is the set of *hbfp null-transitivity constraints* (i.e., $(\forall x,y,z \in C)(Im(f • g) \supset \{(x, y), (y, z)\} \Rightarrow Im(f • g) \supset (x, z) \vee Im(f • g) \supset (x, ) \vee Im(f • g) \supset (, z) \vee Im(f • g) \supset (, ))$.
  - ❖ $\mathcal{HBFPITC}$ is the set of *hbfp intransitivity constraints*, e.g., *Host • Visitor* : *ELIMINATORY_ MATCHES* → *TEAMS* x *TEAMS*.
  - ❖ $\mathcal{HBFPEC}$ is the set of *hbfp Euclideanity constraints*, e.g., *Host • Visitor* : *CHAMPIONSHIP_ MATCHES* → *TEAMS* x *TEAMS* (in (E)MDM, Euclideanity means both left and right Euclideanity).
  - ❖ $\mathcal{HBFPNEC}$ is the set of *hbfp null-Euclideanity constraints* (i.e., $(\forall x,y,z \in C)(Im(f • g) \supset \{(x, y), (y, z)\} \Rightarrow Im(f • g) \supset (y, z) \vee Im(f • g) ) \supset (y, ) \vee Im(f • g) \supset (, z) \vee Im(f • g) \supset (, )) \wedge (Im(f • g) \supset \{(x, y), (z, y)\} \Rightarrow Im(f • g) \supset (x, z) \vee Im(f • g) \supset (x, ) \vee Im(f • g) \supset (, z) \vee Im(f • g) \supset (, ))$.
  - ❖ $\mathcal{HBFPIEC}$ is the set of *hbfp inEuclideanity constraints*, e.g., *Host • Visitor* : *ELIMINATORY_ MATCHES* → *TEAMS* x *TEAMS* (in (E)MDM, inEuclideanity means neither left, nor right Euclideanity).

- ❖ $\mathcal{HBFPQT}$ is the set of *hbfp equivalence constraints*, e.g., TakeOffAirport • LandingAirport : AIRPORT_CONNECTIONS → AIRPORTS x AIRPORTS.
- ❖ $\mathcal{HBFPNQC}$ is the set of *hbfp null-equivalence constraints* (i.e., *f* • *g* is null-equivalent iff it is null-reflexive and null-Euclidean or reflexive and null-Euclidean or null-reflexive and Euclidean).
- ❖ $\mathcal{HBFPACC}$ is the set of *hbfp acyclicity constraints*, e.g., Host • Visitor : ELIMINATORY_MATCHES → TEAMS x TEAMS.

In total, there are 17 hbfp constraint types, out of which only connectivity, reflexivity, null-identity, irreflexivity, symmetry, asymmetry, transitivity, intransitivity, Euclideanity, inEuclideanity, and acyclicity are fundamental: equivalence is derived from reflexivity and Euclideanity, and the other 5 null-type ones are trivially derived from their not-null counterparts.

- ✓ $\mathcal{MSMC} = \mathcal{SMRC} \oplus \mathcal{SMNRC} \oplus \mathcal{SMIRC} \oplus \mathcal{SMSC} \oplus \mathcal{SMNSC} \oplus \mathcal{SMASC} \oplus \mathcal{SMIC} \oplus \mathcal{SMNISC} \oplus \mathcal{SMAIC} \oplus \mathcal{SMRSC} \oplus \mathcal{SMNRSC} \oplus \mathcal{SMAC} \oplus \mathcal{SMQC} \oplus \mathcal{SMNQC}$ is the set of *self-map constraints*, where:
  - ❖ $\mathcal{SMRC}$ is the set of *self-map reflexivity constraints*, e.g., Country ° State ° Capital total, reflex, State ° Capital total.
  - ❖ $\mathcal{SMNRC}$ is the set of *self-map null-reflexivity constraints*, e.g., Country ° State ° Capital reflex, State ° Capital (i.e., Capital might take null values as well).
  - ❖ $\mathcal{SMIRC}$ is the set of *self-map irreflexivity constraints*, e.g., Folder : FILES → FILES, ReportTo : EMPLOYEES → EMPLOYEES, Mother : PEOPLE → PEOPLE, Father : PEOPLE → PEOPLE.
  - ❖ $\mathcal{SMSC}$ is the set of *self-map symmetry constraints*, e.g., Spouse : MARRIED_PEOPLE → MARRIED_PEOPLE total
  - ❖ $\mathcal{SMNSC}$ is the set of *self-map null-symmetry constraints*, e.g., Spouse : PEOPLE → PEOPLE (i.e., Spouse might take null values as well).
  - ❖ $\mathcal{SMASC}$ is the set of *self-map asymmetry constraints*, e.g., Folder : FILES → FILES, ReportTo : EMPLOYEES → EMPLOYEES, Mother : PEOPLE → PEOPLE, Father : PEOPLE → PEOPLE.
  - ❖ $\mathcal{SMIC}$ is the set of *self-map idempotency constraints*, e.g., LocalRepres : CITIZENS → CITIZENS total.
  - ❖ $\mathcal{SMNIC}$ is the set of *self-map null-idempotency constraints*, e.g., ReplacementPart : PART_TYPES → PART_TYPES (i.e., ReplacementPart might take null values as well).
  - ❖ $\mathcal{SMAIC}$ is the set of *self-map anti-idempotency constraints*, e.g., Folder : FILES → FILES, ReportTo : EMPLOYEES → EMPLOYEES, Mother : PEOPLE → PEOPLE.
  - ❖ $\mathcal{SMRSC}$ is the set of *representative system mapping constraints*, e.g., Representative : USCITIZENS → REPRESENTATIVES, Representative = RepresentedBy ° District, where District is the *canonical surjection* District : USCITIZENS → DISTRICTS and RepresentedBy : DISTRICTS ↔ REPRESENTATIVES ⊆ USCITIZENS is the canonical identification mapping of the corresponding representative system.
  - ❖ $\mathcal{SMNCSC}$ is the set of *null-representative system mapping constraints*, e.g., ReportsTo : EMPLOYEES → EMPLOYEES (i.e., ReportsTo might take null values as well, as, generally, the roots of hierarchical organizations do not report to anybody).
  - ❖ $\mathcal{SMQC}$ is the set of *self-map equivalence constraints*, e.g., State ° Capital total.
  - ❖ $\mathcal{SMNQC}$ is the set of *self-map null-equivalence constraints*, e.g., State ° Capital (i.e., Capital might take null values as well).
  - ❖ $\mathcal{SMAC}$ is the set of *self-map acyclicity constraints*, e.g., Folder : FILES → FILES, ReportTo : EMPLOYEES → EMPLOYEES, Mother : PEOPLE → PEOPLE, Father : PEOPLE → PEOPLE.

In total, there are 14 types of self-map constraints, none of which being fundamental, as self-maps are particular cases of dyadic relations (where $\mathbf{1}_D$ and $f : D \to D$ are the canonical projections).

- ✓ $\mathcal{FDCC} = \mathcal{FDEC} \oplus \mathcal{FDNCC} \oplus \mathcal{FDACC} \oplus \mathcal{FDLCC} \oplus \mathcal{FDLNCC} \oplus \mathcal{FDLACC} \oplus \mathcal{FDGCC} \oplus \mathcal{FDLSC} \oplus \mathcal{FDLNSC} \oplus \mathcal{FDLASC} \oplus \mathcal{FDLIC} \oplus \mathcal{FDLNIC} \oplus \mathcal{FDLAIC} \oplus \mathcal{FDLQC} \oplus \mathcal{FDLNQC} \oplus \mathcal{FDLACC} \oplus \mathcal{FDLRSC} \oplus \mathcal{FDLCNRSC}$ is the set of *function diagram cycle constraints*, where:
  - ❖ $\mathcal{FDEC}$ is the set of *function diagram commutativity (equality) constraints*, e.g., PC ° LogicDrive = Host ° DBMS ° DB ("any file which is managed by a DBMS should belong to a logic drive of the PC hosting that DBMS").

- ❖ *FDNCC* is the set of *function diagram null-commutativity constraints*, e.g., *RContinent = MContinent ° Range ° Subrange ° Group ° Mountain* ("any river that springs from a mountain belongs to the same continent as that mountain"; both *Mountain* and *Group* might take null values as well).
- ❖ *FDACC* is the set of *function diagram anti-commutativity (inequality) constraints*, e.g., $(\forall x \in NEIGHBOR\_COUNTRIES)(FrontierColor(Country(x)) \neq FrontierColor(Neighbor(x)))$.
- ❖ *FDLCC* is the set of *function diagram local commutativity (reflexivity) constraints*, e.g., *ConstrRelation ° PrimaryKey* = $\mathbf{1}_{RELATIONS}$ ("the primary key of any relation is a constraint of that relation").
- ❖ *FDLNCC* is the set of *function diagram local null-commutativity constraints*, e.g., *Country ° State ° Capital* = $1_{COUNTRIES}$ ("the capital of a country must be a city of that country"; *Capital* might take null values as well)
- ❖ *FDLACC* is the set of *function diagram local anti-commutativity (irreflexivity) constraints*, e.g., *Spouse ° Mother, Spouse ° Father, Mother ° Spouse, Father ° Spouse.*
- ❖ *FDLSC* is the set of *function diagram local symmetry constraints*, e.g., *Spouse ° * $\mathbf{1}_{PEOPLE}$ total.
- ❖ *FDLNSC* is the set of *function diagram local null-symmetry constraints*, e.g., *Spouse ° * $\mathbf{1}_{PEOPLE}$ (i.e., *Spouse* might take null values as well).
- ❖ *FDLASC* is the set of *function diagram local asymmetry constraints*, e.g., *Spouse ° Mother, Spouse ° Father, Mother ° Spouse, Father ° Spouse.*
- ❖ *FDLIC* is the set of *function diagram local idempotency constraints*, e.g., *Country ° State ° Capital* total.
- ❖ *FDLNIC* is the set of *function diagram local null-idempotency constraints*, e.g., *Country ° State ° Capital* (i.e., *Capital* might take null values as well).
- ❖ *FDLAIC* is the set of *function diagram local anti-idempotency constraints*, e.g., *Spouse ° Mother, Spouse ° Father, Mother ° Spouse, Father ° Spouse.*
- ❖ *FDLQC* is the set of *function diagram local equivalence constraints*, e.g., *State ° Capital* total.
- ❖ *FDLNQC* is the set of *function diagram local null-equivalence constraints*, e.g., *State ° Capital* (i.e., *Capital* might take null values as well).
- ❖ *FDLACC* is the set of *function diagram local acyclicity constraints*, e.g., *Spouse ° Mother, Spouse ° Father, Mother ° Spouse, Father ° Spouse.*
- ❖ *FDLRSC* is the set of *function diagram local representative system mapping constraints*, e.g., *currentMP ° EDistrict : UK_VOTERS → UK_VOTERS*, where *currentMP : ELECTORAL_DISTRICTS → UK_VOTERS* total and *EDistrict : UK_VOTERS → ELECTORAL_DISTRICTS* total.
- ❖ *FDLNRSC* is the set of *function diagram local null-representative system mapping constraints*, e.g., *RepresentedBy = currentMP ° EDistrict : UK_VOTERS → UK_VOTERS*, where *currentMP : ELECTORAL_DISTRICTS → UK_VOTERS* and *EDistrict : UK_VOTERS → ELECTORAL_DISTRICTS* (i.e., both *currentMP* and *EDistrict* might take null values as well).
- ❖ *FDGCC* is the set of *function diagram general commutativity constraints*, e.g., let us consider the function diagram from Figure 1 and its associated general commutativity constraint *retPurCnstr*: $(\forall x \in PUR\_DETAILS)(\forall y \in RET\_DETAILS)(Purchase(x) = Purchase(Return(y)) \land Product(x) = Product(y) \Rightarrow Rqty(y) \leq Pqty(x))$ ("You cannot return more products than you have purchased").

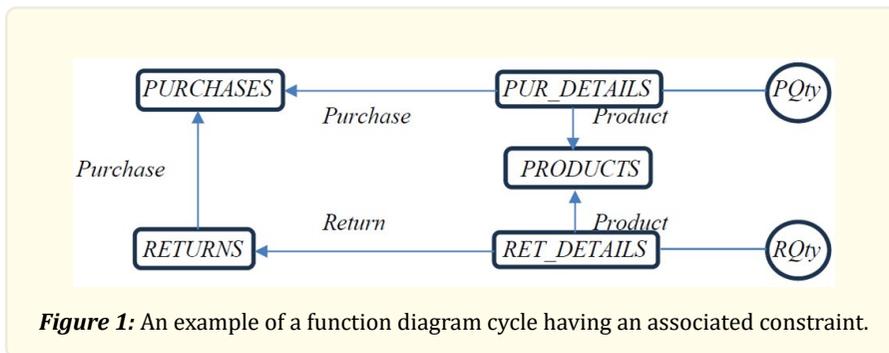

*Figure 1:* An example of a function diagram cycle having an associated constraint.

In total, there are 18 types of function diagram cycle constraints, out of which only the commutativity, general commutativity, and anti-commutativity are fundamental: the local-type ones are particular cases of self-maps, namely computed self-maps through function composition.

The grand total for the mapping constraint set is 59 types, with 22 ones fundamental.

*Object constraints*

In (E)MDM, any other explicit constraint (i.e., which is not of any of the above 75 types) is called an *object constraint* and is formalized by a closed Horn clause. This type is fundamental: in fact, any other constraint type is ultimately formalized by a closed Horn clause.

➢ *OC* is the set of object constraints, e.g., let us consider the function diagram from Figure 2 (where *Im*(*Type*) = {"cash", "card", "wire"}), functions, and associated object (*PD1*) and non-existence (*PD2* and *PD3*) constraints:
PD1: $(\forall x \in PAYM\_DOCS)(PDType(Type(x)) = $ "cash" $\Rightarrow To(x) \in NULLS \land From(x) \in NULLS \land Card(x) \in NULLS)$ ("Whenever Type is "cash", To, From, and Card must be null").
PD2: $To \bullet Card \neg \vdash From$ ("If Type is "card", From must be null, whereas To and Card must not be null").
PD3: $To \bullet From \neg \vdash Card$ ("If Type is "wire", Card must be null and To and From must not be null").

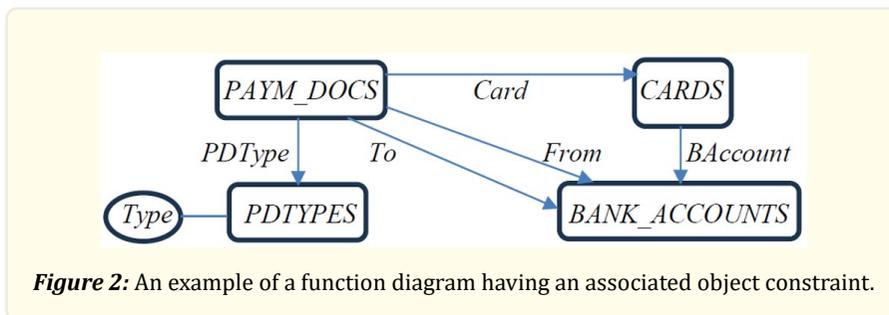

*Figure 2:* An example of a function diagram having an associated object constraint.

In total, (E)MDM provides 76 types of explicit constraints, out of which 24 are fundamental.

The 6 types of relational constraints provided by RDBMSes (range ((co)domain), not null, keys, referential integrity (foreign keys), tuple (check), and default value) should be enforced through their engines. All other constraint types (i.e. non-relational) should be enforced by the software applications managing the corresponding dbs, through their event-driven methods and embedded SQL.

*System constraints*

Apart from the explicit ones, any db scheme also includes implicit *system constraints* (all being transparent to users). Their types and numbers vary in function of the DBMS versions implementing them. The following are among the most common ones:

- $SysC$ is the set of *system constraints*, e.g.,
  - $\varnothing \subseteq S, \forall S \in S.$
  - $card(\varnothing) = 0; card(S) \geq 0, \forall S \in S.$
  - $\varnothing \cup S = S, \varnothing \oplus S = S, \forall S \in S.$
  - $\varnothing \cap S = \varnothing, S \cap S = S, S \cup S = S, \forall S \in S.$
  - $T = S \Leftrightarrow T \subseteq S \wedge S \subseteq T, \forall S, T \in S.$
  - $U = S \oplus T \Leftrightarrow S \cap T = \varnothing \wedge S \cup T = U, \forall S, T, U \in S.$
  - $NAT(n) \subseteq INT(n) \subseteq CURRENCY(n) \subseteq RAT(n + m, m)$, for any naturals $n$ and $m > 1$.
  - $NAT(n) \subset NAT(m)$, $INT(n) \subset INT(m)$, ..., $ASCII(n) \subset ASCII(m)$, for any naturals $n < m$.
  - $\forall R \in \mathcal{R}, R = (f1 \rightarrow C, ..., fn \rightarrow C), f1$ total $\wedge ... \wedge fn$ total $\wedge f1 \bullet ... \bullet fn$ injective.

**An (E)MDM scheme example**

Figure 3 presents, as an example, the (E)MDM scheme for a simple subuniverse of interest: people and marriages (in legislations in which polygamy is not allowed).

**Discussion**

For example, you can test a week for free all following popular tree family software products and, if you use your imagination, you will be amazed by how poor is their concern for your data quality. To exemplify with only a few commonsense tests, please note that, with the products Legacy Family Tree (LFT) [11], RootsMagic (RM) [12], GenoPro (GP) [13], Gramps (G) [14], Family Historian (FH) [15], Family Tree Heritage (FTH) [16], and Ancestral Quest (AQ) [17] you can store in your db values according to which any person was:

- born after his/her death (LFT, RM, GP, G, FH, FTH, AQ).
- baptized before birth (LFT, RM, G, FTH, AQ).
- buried before death (RM, G, FTH, AQ).
- having a male as mother (FH).
- born before his/her parents' birth or hundreds of years after their death (GP, G, FH, FTH, AQ).
- having a same person as both his/her mother and father (G).
- married with a person of same sex and had together a child dead before their birth (G).

It is true that, for some values of such totally unplausible data, FH, FTH, and AQ (a clone of FTH) at least display a warning reading that they might be wrong, but if you ignore it you may save them in the db.

Obviously, thanks to its constraints, the (E)MDM scheme presented in Figure 3 (correspondingly extended with baptism and burial dates, as well as corresponding constraints) would reject storing any such unplausible data.

**Related work**

Besides its system constraints, (E)MDM also includes dozens of meta-constraints for guaranteeing minimality of constraint sets and assisting its users in also guaranteeing their coherence [8]. For example, as acyclicity implies asymmetry, which implies irreflexivity, both asymmetry and irreflexivity are redundant for acyclic graphs (be them dyadic relations or self-maps or homogeneous binary function products), while acyclicity and reflexivity or/and symmetry are incoherent.

**PEOPLE** (The set of world persons of interest.)

    $x \leftrightarrow AUTON(38)$, *total*

    $FName \rightarrow ASCII(32)$, *total* (First names are mandatory.)

    $LName \rightarrow ASCII(128)$, *total* (Last name is mandatory.)

    $Sex \rightarrow \{\text{'F'}, \text{'M'}\}$, *total* (Sex is mandatory.)

    $BirthDate \rightarrow [1/1/1900, Today()]$, *total* (Birth date is mandatory.)

    $Died \rightarrow [1/1/1900, Today()]$

    $SSN \rightarrow NAT(9)$, *total* (SSNs are mandatory.)

$Mother : PEOPLE \rightarrow PEOPLE$, *acyclic*

$Father : PEOPLE \rightarrow PEOPLE$, *acyclic*

$BirthPlace : PEOPLE \rightarrow CITIES$

$HomeTown : PEOPLE \rightarrow CITIES$, *total* (Hometowns are mandatory.)

*$HomeCountry = Country \circ State \circ HomeTown$

$peopleKey: SSN \bullet {}^{*}HomeCountry$ *key* (There may not be two persons of a same country having same SSNs.)

$C_0$: $(\forall x \in PEOPLE)(Died(x) \notin NULLS \Rightarrow BirthDate(x) \leq Died(x))$ (Nobody may die before being born.)

$C_5$: $(\forall x \in PERSONS)(BirthDate(x) + 160 * 365 \geq Died(x) \geq BirthDate(x))$ (nobody may live more than 160 years)

$C_6$: $(\forall x \in PERSONS)(Sex(Mother(x)) = \text{'F'})$ (only women may be mothers)

$C_7$: $(\forall x \in PERSONS)(Sex(Father(x)) = \text{'M'})$ (only men may be fathers)

**Figure 3:** An example of an (E)MDM scheme.

$C_8$: $(\forall x \in PERSONS)(BirthDate(x) \leq Died(Mother(x)) \wedge BirthDate(x) + 5 * 365 \leq BirthDate(Mother(x)) \leq BirthDate(x) + 75 * 365)$ (i.e., no mother may give birth after her death, before being 5 years old, or after being 75 years old)

$C_9$: $(\forall x \in PERSONS)(BirthDate(x) \leq Died(Father(x)) + 10 * 30 \wedge BirthDate(x) + 9 * 365 \leq BirthDate(Father(x)) \leq BirthDate(x) + 100 * 365)$ (i.e., no father may have a child after 10 months from his death, before being 9 years old, or after being 100 years old)

**MARRIAGES** (The set of marriages of interest.)

    $x \leftrightarrow AUTON(38)$, *total*

    $MarriageDate \rightarrow [1/1/1900, Today()]$, *total* (Marriage date is mandatory)

    $DivorceDate \rightarrow [1/1/1900, Today()]$

$Husband : MARRIAGES \rightarrow PEOPLE$, *total*

$Wife : MARRIAGES \rightarrow PEOPLE$, *total*

$C_1$: $(\forall x \in MARRIAGES)(Sex(Wife(x)) = \text{'F'} \wedge Sex(Husband(x)) = \text{'M'})$ (Both wives and husbands should have corresponding sexes.)

$C_2$: $(\forall x \in MARRIAGES)(DivorceDate(x) \notin NULLS \Rightarrow MarriageDate(x) < DivorceDate(x))$ (Nobody can divorce before being married.)

$C_3$: $(\forall x \in MARRIAGES)(MarriageDate(x) \geq BirthDate(Wife(x)) + 16 * 365 \wedge MarriageDate(x) \geq BirthDate(Husband(x)) + 16 * 365 \wedge (Died(Wife(x)) \in NULLS \vee MarriageDate(x) < Died(Wife(x))) \wedge (Died(Husband(x)) \in NULLS \vee MarriageDate(x) < Died(Husband(x))))$ (Nobody can get married before being 16 years old or after dying.)

**Figure 3:** (Continued).

$C_4$: $(\forall x \in MARRIAGES)(DivorceDate(x) \notin NULLS \Rightarrow (Died(Wife(x)) \notin NULLS \Rightarrow DivorceDate(x) < Died(Wife(x))) \wedge (Died(Husband(x)) \notin NULLS \Rightarrow DivorceDate(x) < Died(Husband(x))))$ (Nobody may divorce after dying.)

$C_{10}$: $Husband \bullet MarriageDate$ key (nobody may marry same day more than once)

$C_{11}$: $Wife \bullet MarriageDate$ key (nobody may marry same day more than once)

$C_{12}$: $Husband \bullet DivorceDate$ key (nobody may divorce same day more than once)

$C_{13}$: $Wife \bullet DivorceDate$ key (nobody may divorce same day more than once)

$C_{14}$: $(\forall x,y \in MARRIAGES, x \neq y)(Husband(x) = Husband(y) \Rightarrow DivorceDate(x) \notin NULLS \wedge MarriageDate(y) > DivorceDate(x) \vee DivorceDate(y) \notin NULLS \wedge MarriageDate(x) > DivorceDate(y))$ (polygamy is not allowed: one can remarry only after divorcing)

$C_{15}$: $(\forall x,y \in MARRIAGES, x \neq y)(Wife(x) = Wife(y) \Rightarrow DivorceDate(x) \notin NULLS \wedge MarriageDate(y) > DivorceDate(x) \vee DivorceDate(y) \notin NULLS \wedge MarriageDate(x) > DivorceDate(y))$ (polygamy is not allowed: one can remarry only after divorcing)

$C_{16}$: $(\forall x \in MARRIAGES)((Died(Husband(x)) > MarriageDate(x) \vee Died(Husband(x)) \in NULLS) \wedge MarriageDate(x) > 16 * 365 + BirthDate(Husband(x)))$ (nobody can marry before being 16 or after his death)

$C_{17}$: $(\forall x \in MARRIAGES)((Died(Wife(x)) > MarriageDate(x) \vee Died(Wife(x)) \in NULLS) \wedge MarriageDate(x) > 16 * 365 + BirthDate(Wife(x)))$ (nobody can marry before being 16 or after her death)

$C_{18}$: $(\forall x \in MARRIAGES)(Died(Husband(x)) > DivorceDate(x) \vee Died(Husband(x)) \in NULLS)$ (nobody can divorce after his death)

$C_{19}$: $(\forall x \in MARRIAGES)(Died(Wife(x)) > DivorceDate(x) \vee Died(Wife(x)) \in NULLS)$ (nobody can divorce after her death)

***Figure 3:*** *(Continued).*

We discussed the extraordinary importance of db constraints for guaranteeing data quality in [18, 19].

We used the (E)MDM to model the RDM [6] and the first order predicate logic with equality [20].

*MatBase* is our intelligent data and knowledge management system prototype, based on both (E)MDM, E-RDM, and RDM [21-23].

We published several algorithms for detecting some of the (E)MDM constraint types [24-27], as well as for enforcing them in *MatBase* [26-32]. Papers [33, 34] deal with the implementation and usage, respectively, of the *MatBase's* Datalog¬ subsystem.

We proved that the expressive powers of E-RDM [10, 35, 36] and the Functional Data Model (FDM) [37] (which was the data model that inspired (E)MDM are equivalent [38].

Apart from the FDM, other data modeling approaches are related to (E)MDM: categorical [39], graph [40-43], incomplete dbs [44], probabilistic [45], possibilistic [46], and constraint dbs [47].

Finally, the most closely related approaches to non-relational constraint enforcement are based on business rules management (BRM) [48, 49] and their corresponding implemented systems (BRMS) and process managers (BPM), like the IBM Operational Decision Manager [50], IBM Business Process Manager [51], Red Hat Decision Manager [52], Agiloft Custom Workflow/BPM [53], etc. They are generally based on XML (but also on the Z notation, Business Process Execution Language, Business Process Modeling Notation, Decision Model and Notation, or the Semantics of Business Vocabulary and Business Rules), which is the only other field of endeavor trying to systematically deal with business rules, even if informally, not at the db design level but at the software application one, and without providing automatic code generation.

From this perspective, (E)MDM is also a BRM but a formal one, and *MatBase* is also a BRMS but an automatically code generating one.

Data quality may be further improved only semantically, for example by checking values against public and/or private trusted dbs, like Geographic Data [54], Top Companies Database [55], Yellow Pages Scraper [56], Amazon Product Database [57].

## Conclusion

We revisited (E)MDM, including all its 76 current constraint types, which are its main strength in guaranteeing db data plausibility, the highest possible level of syntactical data quality.

We provided lot of real-life examples for all (E)MDM set, mapping, and constraint types.

We also provided clues on implementing (E)MDM schemata in RDBMSes and software applications managing their dbs.

We presented an (E)MDM scheme for a family subuniverse and compared it with some popular family tree software products, proving that it would reject any of the absurd unplausible data value that they accept for storing in their dbs.

We provided a comprehensive batch of related work, illustrated with a rich corresponding reference list, even if it contains only a few seminal items for any other related data modeling approach or DBMS.

## Conflict of Interest

The authors declare that the research was conducted in the absence of any commercial or financial relationships that could be construed as a potential conflict of interest.

## Funding

This research received no external funding.

## Acknowledgments

This research was not sponsored by anybody and nobody other than its author contributed to it.